\documentclass[twocolumn]{aastex62}

\usepackage{textcomp}
\usepackage{amsmath}

\received{2018 November 01}
\revised{2019 February 08}
\accepted{2019 February 15}

\submitjournal{The Astrophysical Journal}

\begin{document}

\title{Sticking Properties of Silicates in Planetesimal Formation Revisited}

\author[0000-0001-5375-2618]{Tobias Steinpilz}
\author{Jens Teiser}
\author{Gerhard Wurm}
\altaffiliation{University of Duisburg-Essen, Faculty of Physics, \newline Lotharstr. 1-21, Duisburg 47057, Germany}

\correspondingauthor{Tobias Steinpilz}
\email{tobias.steinpilz@uni-due.de}

\begin{abstract}
In the past, laboratory experiments and theoretical calculations showed a mismatch in derived sticking properties of silicates in the context of planetesimal formation. It has been proposed by \citet{Kimura2015} that this mismatch is due to the value of the surface energy assumed, supposedly correlated to the presence or lack of water layers of different thickness on a grain's surface. We present tensile strength measurements of dust aggregates with different water content here. The results are in support of the suggestion by \citet{Kimura2015}. Dry samples show increased strengths by a factor of up to 10 over wet samples. A high value of $\gamma = 0.2$\,J/m$^2$ likely applies to the dry low pressure conditions of protoplanetary disks and should be used in the future.
\end{abstract}

\keywords{methods: miscellaneous, planets and satellites: formation, planets and
satellites: fundamental parameters, protoplanetary disks}

\section{Introduction}
Surface energy is an important parameter that determines the sticking of dust particles during planetesimal formation. \citet{Kimura2015} discussed the specific value of surface energy of amorphous silica in great detail. They did a comprehensive literature search finding orders of magnitude differences between values of the surface energy deduced by varying authors. As essence for the variation, they pin down the water content of the surface. Warm or under the conditions of vacuum, the water content is low and silanol groups dominate surface forces. These are strong. Under normal atmospheric conditions, several layers of water are present. Surface forces are dominated by water interactions, which are weaker. This way \citet{Kimura2015} deduce surface energies of a few times $0.01$\,J/m$^2$ for the wet case and $\gamma \sim 0.2$\,J/m$^2$ for the dry case, a large difference of up to an order of magnitude.

This is important in the context of modeling planetesimal formation. \citet{Dominik1997} used $0.025$\,J/m$^2$ for quartz determined by \citet{Kendall}. Based on their model the sticking velocities  below which a quartz grain of $1.2$\,$\mu$m diameter should stick to a wall should only be $0.09$\,m/s. However, \citet{Poppe2000} find about 1 m/s sticking velocity. This is an order of magnitude larger than calculated. Similarly high values also show up in other laboratory experiments e.g. by \citet{Blum2000}.
In any case, the sticking velocity scales with $\gamma^{5/6}$ \citep{Dominik1997}. Thus, the mismatch could be turned into agreement if the surface energy would be higher by more or less a factor of 10, or $\gamma \sim 0.2$\,J/m$^2$ \citep{Kimura2015}.

So the question is, can the lack of water at the protoplanetary disk's low pressure conditions lead to such high values as suggested by \citet{Kimura2015}? In a number of experiments we measured the tensile strength of dust aggregates composed of dry or wet grains to answer this question.

\section{Experiment}

\begin{figure}[ht]
 	\includegraphics[width=\columnwidth]{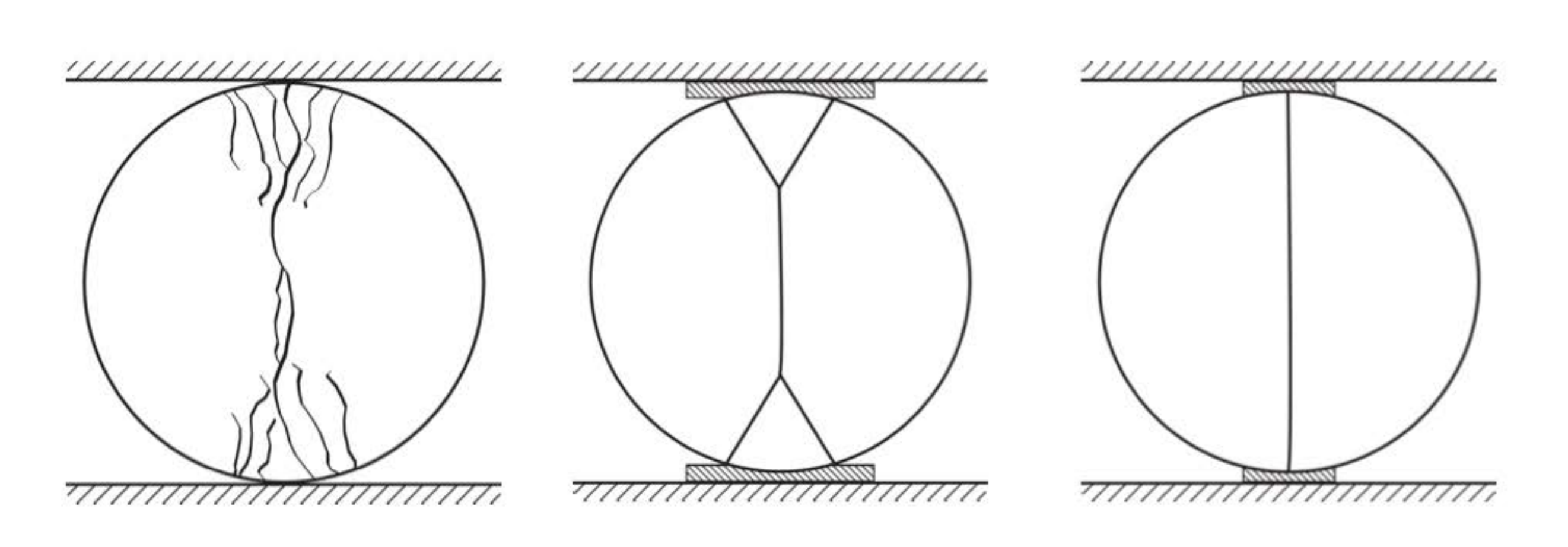}
	\caption{\label{fig.setup}Brazilian test to measure the tensile strength. A force is applied on the top until the cylindrical aggregates break. Image adaped from \citet{malarics} showing on the left and in the middle the example of inappropriately applied forces and the expected crack for a valid tensile strength measurement on the right.}
\end{figure}

The basic idea behind the tensile strength measurements is the Brazilian test sketched in fig.\,\ref{fig.setup}.

\begin{figure}[ht]
 	\includegraphics[width=\columnwidth]{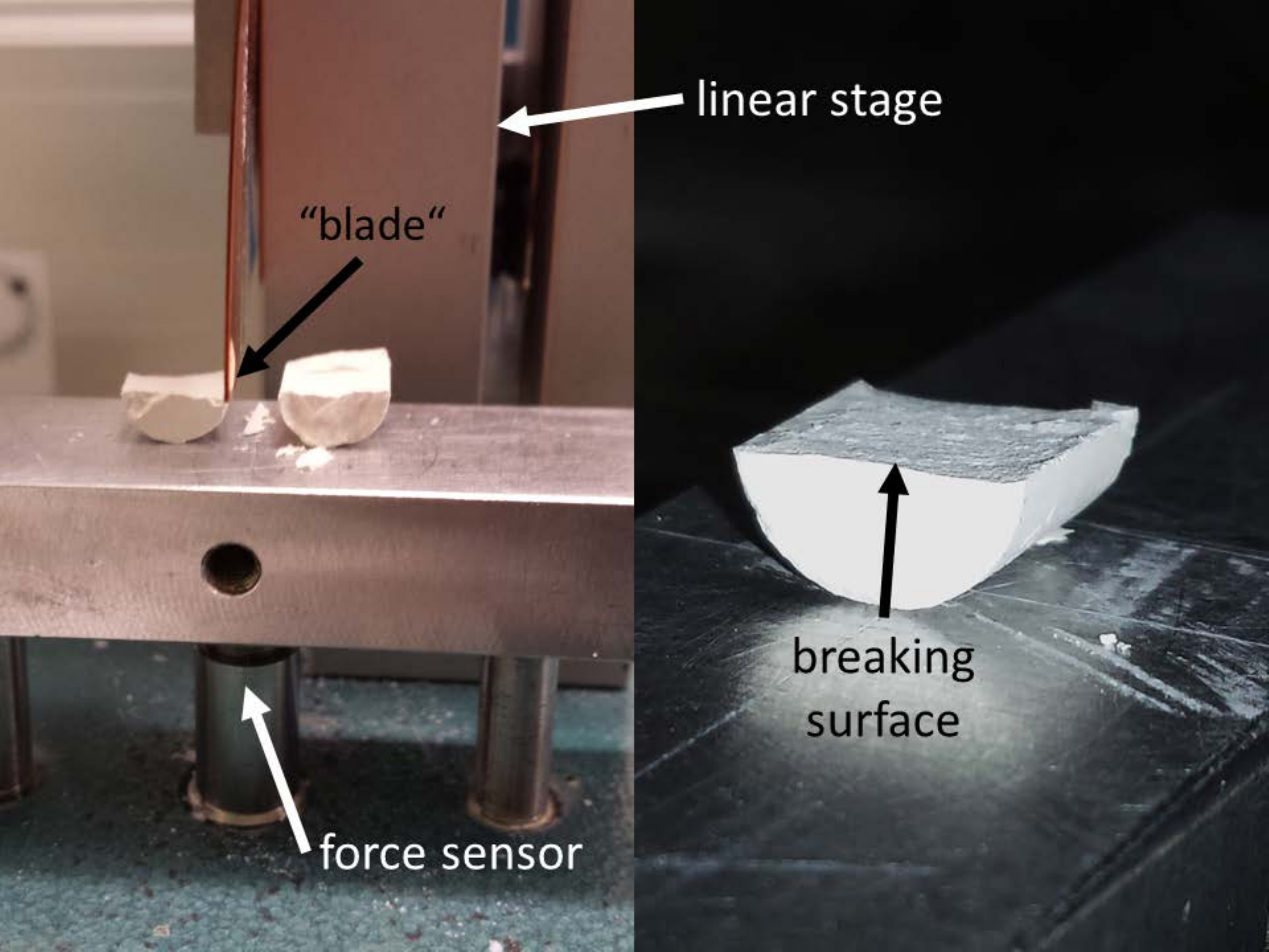}
	\caption{\label{fig.realsetup}\textbf{left}: The experimental setup we used - a force is applied by driving the linear stage via a "blade" onto the cylinder until it breaks in half - via a force sensor the peak force is measured. \textbf{right}: The clean breaking surface of a correctly measured cylinder.}
\end{figure}

This procedure was used before for dust by \citet{Meisner2012} or recently for dust and ice aggregates by \citet{Gundlach2018}. The measured force $F$ at which the aggregate breaks can be translated into a tensile strength $\sigma = 2F/(\pi d L)$ - in our experiment with the aggregate diameter $d \approx 8$\,mm and cylinder length $L =5-9$\,mm. Fig.\,\ref{fig.realsetup} shows the experimental setup we used including a half of a correctly broken cylinder. The experiment is carried out at normal atmosphere but with samples prepared in two different ways.

\section{Samples}
We used a commercial amorphous silica sample by Micromod (Sicastar plain) with a grain size of $1.2$\,$\mu$m. This sample is monodisperse and grains are spherical. They match the grains used in the earlier collision experiments referred to by \citet{Kimura2015}. 

\subsection{Normal atmosphere}
One sample was not treated in any way. That means the dust was just taken from its bottle and pressed into a cylinder with varying filling factor $\Phi$ (volume fraction filled with material). The typical water content of these samples was determined by weighing samples before and after heating to 250\,$^\circ$C for 24\,h. This shows a water content (mass fraction) of the sample of 6.6\,$\pm$\,1.1\,\%, resulting in a corresponding uncertainty of $\Phi$. The surface water content corresponds to a homogeneous 25.3\,$\pm$\,4\,nm thick layer or about 84\,$\pm$\,14\,mono layers of water on the surface of each grain. The volume filling factor was corrected to the pure silica case to achieve a comparable factor. 

\subsection{Heated samples}

\begin{figure}[ht]
 	\includegraphics[width=\columnwidth]{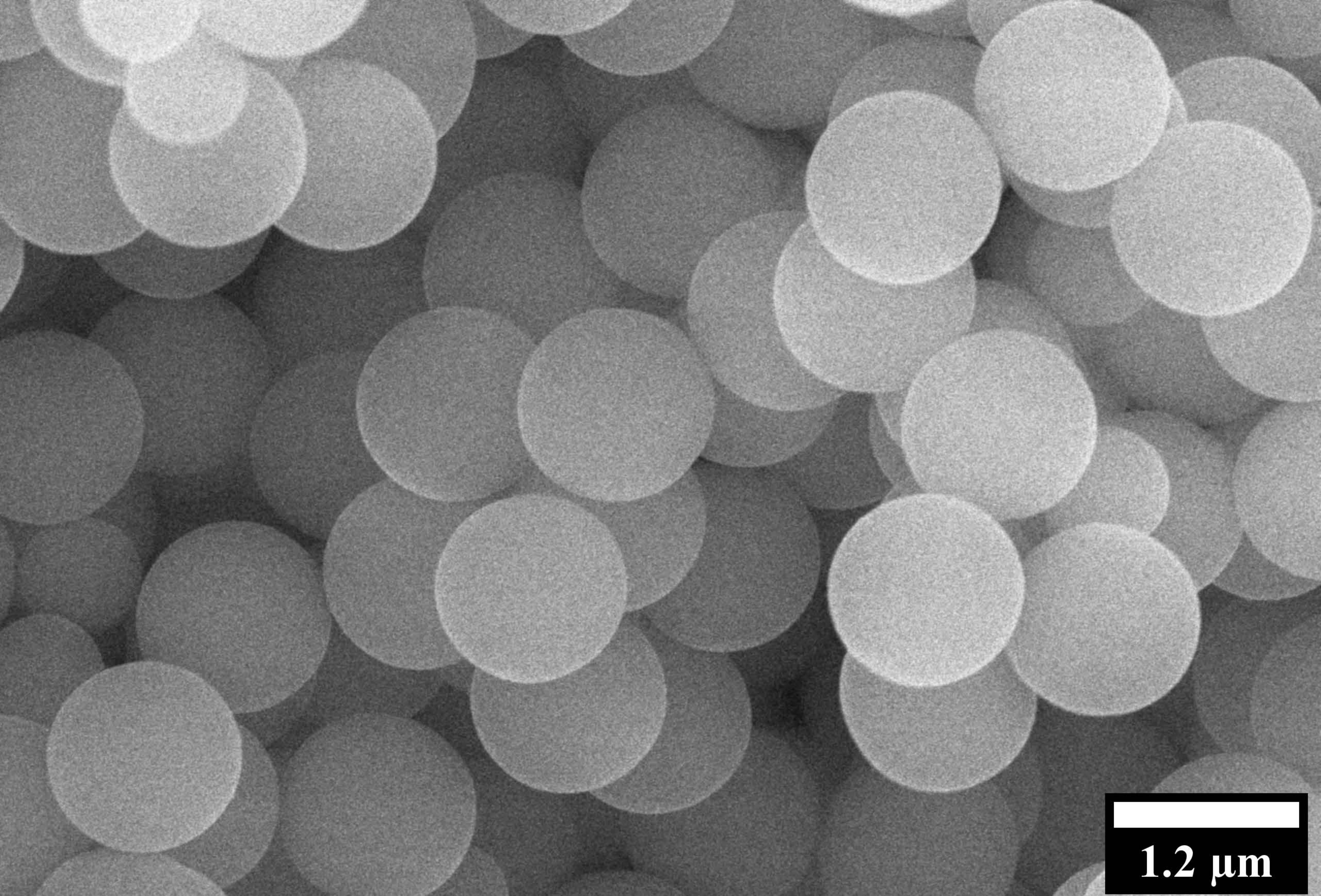}
	\caption{\label{fig.sinter}SEM (15kV) image of a heated sample (24\,h at 300\,$^\circ$C) showing no sintering.}
\end{figure}

The second kind of sample was the heated samples to remove surface water. Measurements were carried out immediately after the sample was taken out of the oven. About 5 samples were heated at a time and measured in sequence. A whole sequence took about 10\,minutes.  During the measurements the samples can recollect water but due to the compact nature, the diffusion times into the dust aggregates are much longer than the measurement times and we consider these aggregates to be dry. The debris was used again for the preparation of new samples. To do so, the fragments were crushed in a mortar and then again pressed to cylindrical agglomerates and in this case of dry samples finally heated to 250\,$^\circ$C for 24\,h. To rule out sintering during the heating procedure we heated a sample for 24\,h at 300\,$^\circ$C and afterwards observed no sinter-necks via SEM (fig.\,\ref{fig.sinter}).

\section{Results}
 
\begin{figure}[h]
	\includegraphics[width=\columnwidth]{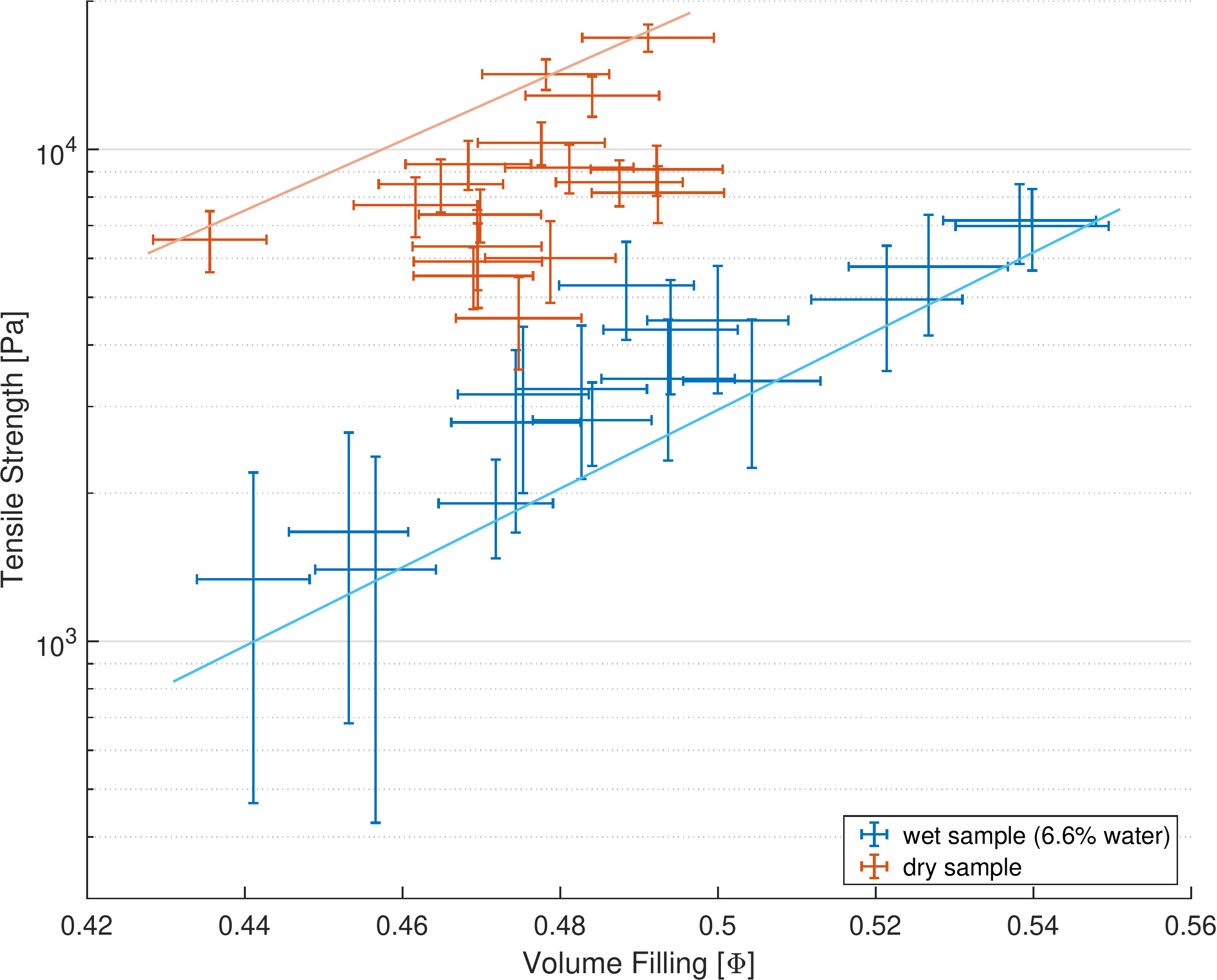}
	\caption{\label{fig.comparesica}Tensile strengths of dry and wet amorphous silica aggregates in comparison for different filling factors. The lines are power laws manually placed to the lowest and highest values.}
\end{figure}

Fig.\,\ref{fig.comparesica}  shows the comparisons of the tensile strength measurements for wet and dry samples.

The wet samples show a clear power law dependence on filling factor as known before \citep{Meisner2012}.
The dry samples are somewhat less constrained with more variation. However, the data of both subsets differ by an order of magnitude. As tensile strength directly depends on the sticking of individual contacts and as aggregates only differ in water content, this directly translates in a difference in surface energy according to Rumpf's equation \citep{Rumpf1970}
\begin{equation}
\label{rumpf}
\sigma = \frac{9 \Phi N}{8 \pi d^2} (f_c \gamma) 
\end{equation}
Here, N is the coordination number which is on the order of 5. The contact force inserted is $F_c = f_c \gamma$ as all forces of importance are proportional to $\gamma$. This includes the pull-off force, the rolling force and essentially also the sliding force. We therefore conclude that dry samples can have a factor 10 higher surface energy than wet samples. In more detail \citet{Omura2017} find that the contact force entering for very cohesive dust is somewhere between rolling force and sliding force depending on the porosity. If the coordination number is low more particles can roll. For our dense aggregates, the relevant force should be closer to sliding. For the ratio between sliding and rolling force for silica beads of similar size \citet{Omura2017} calculate a factor of about 100. Putting in a rolling force of $F_r = 6 \pi \gamma \xi$ with a critical displacement of 0.2 nm \citep{Dominik1997} the factor $f_c$, if we assume sliding, is about $f_c = 0.12 \pi \rm \mu m$.
If we put this into eq. \ref{rumpf} we get $\gamma=0.013$\,J/m$^2$ for the wet case based on pure sliding. The real value will be somewhat higher as on average some particles will still be allowed to roll. Details are beyond this work but this estimate is in agreement to the literature values compiled by \citet{Kimura2015}

We attribute the strong variation in the dry samples compared to the wet samples to the preparation process. The data show higher tensile strengths for dry samples, which directly means that also the resulting fragments after an experiment are more stable in the dry case. As these fragments are crushed and pressed into a cylindrical shape for following experiments, some sub-structure within the pressed agglomerate might remain after the preparation process and weaken the resulting agglomerates. This explanation most likely applies for the extremely low values around $\Phi \sim 0.48$ in fig.\,\ref{fig.comparesica} since these points where created at the end of our measurements so out of the most often recycled material. Similar effects have been observed by \citet{Schraepler2012}. 

We did not analyze the structure of the aggregates further or optimized the preparation process as we think the data already make the point sufficiently clear that surface water changes the surface energy strongly for the silica grains. \citet{Kimura2015} actually also consider some variation or two values $\gamma \sim 0.15$\,J/m$^2$ and  $\gamma \sim 0.25$\,J/m$^2$ for specific experiment cases as best match. Our experiments show that such high values might indeed be justified.

\section{Conclusion}

We find that the water content on the surface of silica grains has a significant influence on the value of the surface energy. We compared the surface energy of dry and wet samples. They can differ up to a factor of 10. We did not directly measure values under vacuum or under low pressure but following the arguments by \citet{Kimura2015} these should be similar to the values of the dry samples. Our findings are therefore supporting the suggestion by \citet{Kimura2015} that choosing a high value for the surface energy on the order of $\gamma \sim 0.2$\,J/m$^2$ would be appropriate to match calculations to the experiments.  As conditions in protoplanetary disks also match the dry cases of the experiments, conclusions drawn from the specific experiments so far still apply, i.e. that the sticking threshold of micrometer grains is on the order of 1 m/s. 

In any case, following \citet{Kimura2015} we propose the use of high values for the surface energy for numerical simulation on dust growth of silicates to obtain consistent and protoplanetary disks like results.

\end{document}